**Article type:** Full Paper

**Formation and Stability of Ordered and Disordered Ba-substituted Calcite Phases**

*Eva Seknazi, Davide Levy†, Iryna Polishchuk, Alex Katsman, Boaz Pokroy\**

E. Seknazi, Dr. D. Levy, Dr. I. Polishchuk, Prof. A. Katsman, Prof B. Pokroy

Department of Materials Science and Engineering and the Russel Berrie Nanotechnology Institute, Technion - Israel Institute of Technology, Haifa 32000, Israel

E-mail: bpokroy@tx.technion.ac.il

Calcite has the ability to host large amounts of intracrystalline inclusions, a phenomenon that is known to be the case in biominerals and has been demonstrated in bio-inspired synthetic systems. In this study, we focused on barium as the inclusion. Highly substituted Ba-calcite possesses disordered carbonate orientations, characteristic of $R\bar{3}m$ symmetry. We show that calcite undergoes an ordered-disordered transition, in which the carbonate groups undergo progressive loss of their rotational order with increasing amounts of incorporated Ba, and reach complete rotational disorder for a critical amount of Ba. We characterize this transition and propose a theoretical model justifying the influence of Ba. Moreover, the disordered $R\bar{3}m$ Ba-substituted calcite has been previously identified as a high-temperature phase or as a highly metastable room-temperature phase. We challenge those descriptions by successfully synthesizing it under slow-rate conditions and by studying its thermal behavior, and we conclude that the fully disordered Ba-calcite is stable, whereas the transitional, partially disordered Ba-calcite phase is metastable

# 1. Introduction

Many biominerals demonstrate excellent properties associated with highly controlled and elaborated compositions and structures.[1-9] Most marine biominerals consist of calcite, which incorporates intracrystalline organic[1-4] and inorganic inclusions[5-7] that are believed to be important for their superior properties.[1-2, 8-9] Inspired by the strategies of biominerals, and in attempts to gain insights into their mechanisms, numerous studies have addressed the effects and the limits of organic or inorganic intracrystalline inclusions in synthetic calcite.[10-13] Moreover, biogenic calcite is formed via crystallization of an amorphous precursor, amorphous calcium carbonate (ACC).[14-16] This non-classical crystallization pathway is believed to be the key to the formation of the complex composites that constitute biogenic calcite. Indeed, the use of an amorphous precursor facilitates the incorporation of inclusions, which then remain entrapped in the calcium carbonate phase during and after its crystallization. Accordingly, numerous bio-inspired studies have addressed the potential utility of an amorphous precursor, using ACC as the precursor for synthesizing calcite with a high level of inclusion incorporation.[17-23] That is how Mg-calcite, whose formation is kinetically and thermodynamically disfavored,[24] was successfully synthesized from crystallization of Mg-rich ACC (Mg-ACC).[17-21]

Incorporation of Ba in calcite by going through a Ba-rich ACC (Ba-ACC) was studied by Whittaker et al.[23, 25] They showed that direct crystallization of Ba-ACC occurs in water at ambient conditions, and leads to a disordered, highly substituted Ba-calcite phase. This disordered phase belongs to the $R\bar{3}m$ space group and differs crystallographically from calcite ($R\bar{3}c$ space group). The disordered Ba-calcite phase was previously identified at high temperatures,[26] and previously synthesized but incorrectly indexed as barytocalcite.[27] Whittaker et al.,[23] the first to report the formation at ambient conditions of the disordered Ba-calcite phase, named this phase balcite. However, balcite was described as a highly metastable phase, whose synthesis is made possible by the use of the unstable Ba-ACC as a precursor,[23] a claim that we challenge here.

$R\bar{3}m$ symmetry is similar to calcite symmetry ($R\bar{3}c$) but has an additional disorder in the orientations of carbonates about the c-axis. Both symmetries contain layers of carbonates normal to the c-axis. Whereas in $R\bar{3}c$ symmetry the carbonate groups of two consecutive layers are inversely oriented to each other, resulting in an alternating order, the carbonate groups in the $R\bar{3}m$ symmetry occupy one of the two possible orientations randomly, so that the layers are equivalent and the c-parameter of the lattice is halved ($c_{R\bar{3}m} = c_{R\bar{3}c}/2$). This loss of order of the carbonate groups can be detected in X-ray diffractograms with the disappearance of the (hkl) reflections where l is odd. Interestingly, pure calcite undergoes a reversible $R\bar{3}c$ to $R\bar{3}m$ phase transition at high temperature (1240K),[28-31] and the presence of Ba stabilizes this phase down to room temperature. The high-temperature $R\bar{3}m$ calcite phase is called phase V of calcite. Calcite transforms to phase V through an intermediary phase called phase IV (985−1240K), which belongs to the $R\bar{3}c$ space group but whose carbonate groups are partially disordered.[29-31] Phase IV corresponds to the temperature range at which the (hkl) reflections with l odd are decreasing in intensity before disappearing completely.[29-31]

In this study we investigated the incorporation of Ba into calcite using different routes of synthesis and various compositions. We first characterized experimentally the $R\bar{3}c$ to $R\bar{3}m$ transition that occurs with the increase in amount of Ba incorporated in the calcite lattice. We then investigated the relative stability of the ordered and disordered Ba-calcite phases by using slow rate synthesis conditions and by studying their thermal behaviour. Finally, with the help of a theoretical model, we proposed an explanation of the order-disorder transition based on the calcite lattice parameters' changes upon Ba incorporation.

## 2. Results and discussion

## 2.1. $R\bar{3}c$ to $R\bar{3}m$ Phase Transition – experimental results

We chose to study the transition of the ordered to the disordered phase by using powders obtained from hydrothermal synthesis, in which Ba-ACC was used as an amorphous precursor crystallized under conditions of high temperature and pressure (see Experimental Section). Full X-ray diffractograms of the powders obtained from solutions containing various amounts of Ba indicated that the powders were Ba-calcite phases (**Figure 1**a). The increasing Ba content in the solutions used for the syntheses caused shifts of the whole diffractogram towards lower angles (Figure 1b) and a gradual decrease of the (113) peak intensity (Figure 1c). The shift of the diffractogram indicates lattice expansion. Since the ionic radius of Ba (1.35 Å) is larger than that of Ca (1.00 Å),[32] the lattice expansion signals that Ba has become incorporated into the calcite lattice. The observed decrease in (113) peak intensity indicated a gradual disorder in anion orientation about the c-axis. At a critical Ba concentration, the peak disappeared, indicating randomness in the $CO_3$-group orientation corresponding to the $R\bar{3}m$ phase. The partially disordered transitional state was similar to the phase IV of calcite (985−1240K) and the fully disordered state was similar to the calcite phase V (above 1240K). The critical concentration at which Ba-calcite became fully disordered was achieved with the powder synthesized with 25% Ba in solution (i.e. [$BaCl_2$]/([$CaCl_2$]+[$BaCl_2$])=25%). According to Reitveld refinements, this powder was composed of the two phases, $R\bar{3}c$ and $R\bar{3}m$.

The amounts of Ba incorporated in the crystals were quantified both from the X-Ray diffraction (XRD) data via Reitveld refinements and from Inductively Coupled Plasma (ICP) measurements (see Experimental Section). For the fully disordered Ba-calcite ($R\bar{3}m$), the results from ICP were consistent with those obtained from Reitveld refinements (Figure 1d). The apparent discrepancy in the case of partially disordered Ba-calcite can be attributed to the fact that the synthesis yields multiple coexisting Ba-calcite phases when solutions of low Ba content are used. These include the main phase as well as Ba-calcite phases with lower Ba content (not shown in Figure 1d). It can indeed be seen in Figure 1b and c that the peaks with low Ba content were asymmetric. Since examples of non-homogeneous distribution of impurity are profuse in nature,[6, 9] we examined whether the distribution of Ba was indeed homogeneous. For this purpose, we used Scanning Transmission Electron Microscope – Energy Dispersive Spectroscopy (STEM-EDS) to examine a Focused Ion Beam (FIB)-cut cross-section of a Ba-calcite crystal. The acquired high-magnification map indeed verified the homogeneous distribution of Ba (Figure 1d, inset).

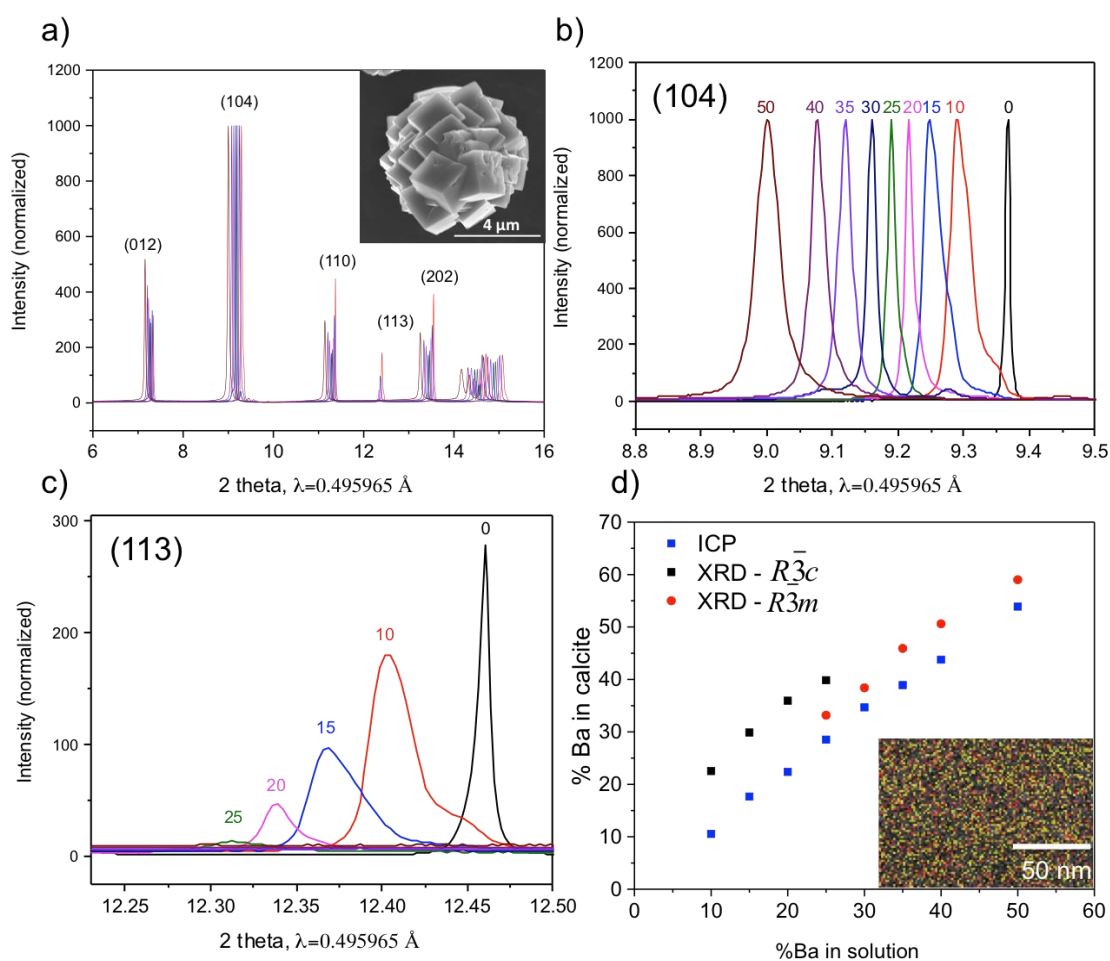

**Figure 1.** a) Full diffractogram (b) (104) and (c) (113) calcite reflections obtained from Ba-calcite powders prepared by hydrothermal synthesis from solutions in which [BaCl2]/([CaCl2]+[BaCl2]) = 0, 10, 15, 20, 25, 30, 35, 40 and 50 %. The peaks were normalized to the (104) peak. The inset in (a) is a High-Resolution Scanning Electron Microscope (HRSEM) image of Ba-calcite synthesized from 50% Ba solutions. (d) Amount of Ba incorporated into calcite as a function of the amount of Ba in the solution used for the synthesis. The inset is a STEM-EDS map of a FIB-cut cross section of Ba-calcite crystals synthesized from a 50% Ba solution. Red and yellow points represent counts from Ca and Ba atoms respectively.

According to refinements from the XRD data, calcite displayed lattice parameter expansions as high as 2.4% and 5.8% for the a- and c-parameters, respectively (**Figure 2**). These lattice distortions were unusually large. In comparison, calcite precipitated with organic molecules experiences lattice expansion due to incorporated molecules with a maximum of 0.5% distortions in the c-parameter.[33] In our study, the critical lattice parameters above which the fully disordered state was achieved were 5.032 ± 0.02 Å and 17.564 ± 0.014 Å for the a- and c-parameters, respectively. Interestingly, the increase in lattice parameters with the amount of incorporated Ba was discontinuous at the order-disorder transition, and the same amount of incorporated Ba caused more distortions when the phase was disordered, indicating that the disordered lattice is inherently larger than the ordered one.

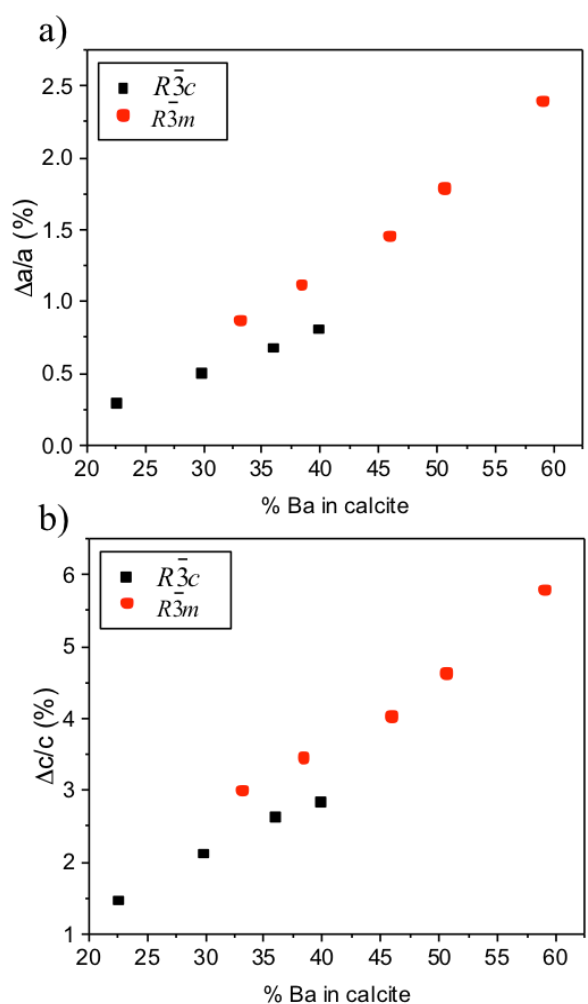

**Figure 2**. Lattice strains (relative to pure calcite) of Ba-calcite powders as a function of the amount of incorporated Ba was calculated from XRD data: (a) distortions in a-parameter, (b) distortions in c-parameter; black, in Ba-calcite (R$\bar{3}$c); red, in balcite (R$\bar{3}$m). The c-parameter of the R$\bar{3}$m phase was multiplied by 2 for continuity with the c-parameter of the R$\bar{3}$c phase.

## 2.2. Incorporation of Ba into Calcite using Slow Synthesis Routes

Balcite (disordered Ba-substituted calcite) synthesis was recently reported using Ba-ACC as a precursor, which was suggested by the authors as a requirement for balcite formation.[23] To determine whether the amorphous precursor is indeed needed for this purpose, or whether balcite can also be formed by slow rates of synthesis, we performed two types of synthesis. For the first route, we used the vapor diffusion technique, in which solutions of $BaCl_2$ and $CaCl_2$, both at low concentrations, were placed for a few days in a sealed desiccator in the presence of $CO_2$ vapor. For the second route we used the drop-by-drop addition technique, in which solutions of $BaCl_2$ and $CaCl_2$, both at low concentrations, were added very slowly (0.05 mL/min) to a solution of $Na_2CO_3$. In both routes we used 10-mM solutions; at lower concentrations no precipitation occurred. Both methods of synthesis resulted in precipitation after a few hours. Thus, these two routes represent slow-rate pathways that favour the formation of thermodynamically stable phases. On the contrary, the hydrothermal synthesis is a fast-rate pathway and the use of the amorphous precursor Ba-ACC may allow the formation of thermodynamically unstable phases.[23, 25]

**Figure 3** shows diffractograms of the powders obtained by these routes of synthesis and with solutions containing 20% and 30% Ba. In all synthesis routes, the (113) peak of calcite was absent when the solution contained 30% Ba, indicating that the phase formed was the disordered balcite ($R\bar{3}m$). Thus, balcite can be formed under both fast-rate and slow-rate conditions. These results are especially interesting since balcite was thought to be a metastable phase at room temperature, and reachable via an amorphous precursor only.[23] Although we did not ensure that there was no ACC involvement in the slow syntheses, the very slow rates and low concentrations used here favour a classical crystallization pathway. With 20% Ba solutions, the slow vapor-diffusion and drop-by-drop syntheses yielded a mix of $R\bar{3}c$ and $R\bar{3}m$ phases. The fact that the $R\bar{3}m$ phase could be formed through slow synthesis but not from an amorphous precursor (for the same fraction of Ba in solution) indicated that the fully disordered phase is preferentially formed under thermodynamic conditions whereas the partially disordered phase preferentially forms under kinetic conditions.

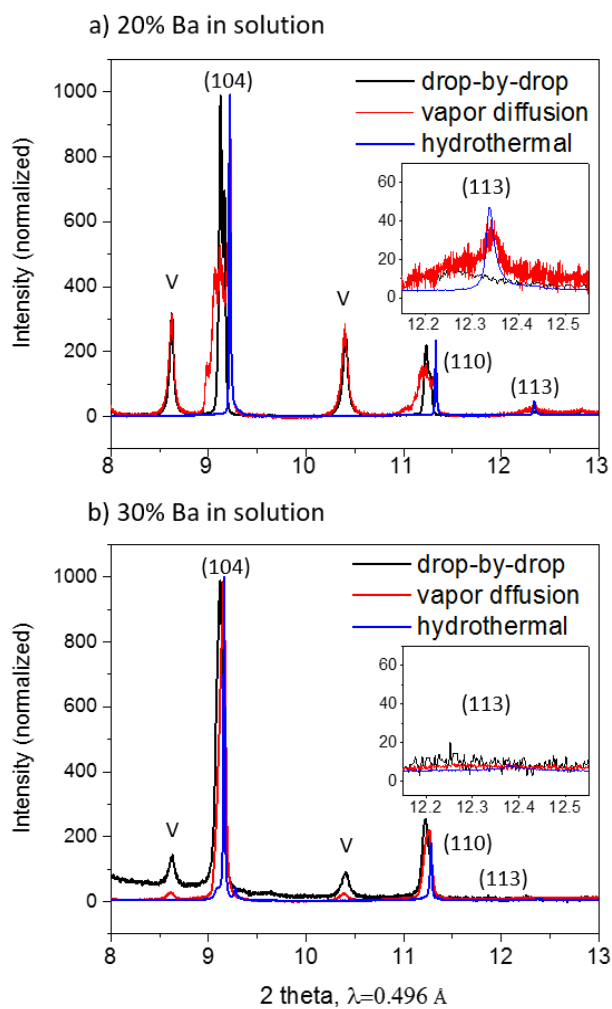

**Figure 3**. Diffractograms of powders obtained using a) 20% Ba, b) 30% Ba solutions and using 3 different routes of syn-thesis: drop-by-drop addition, vapor diffusion and hydro-thermal synthesis. The insets are zoom-ins of the (113) peak position. (V, vaterite)

These results can be compared to the widely studied case of Mg incorporation into calcite, which does not happen at such slow rates because it is limited by two factors: Mg is strongly hydrated,[34] causing Mg ions to dehydrate only partially when attaching calcite nuclei, thus hindering its growth; and the large difference in ionic size between Mg and Ca (Ca, 1.00 Å; Mg, 0.72 Å),[32] which would cause unfavourable high lattice distortions.[19, 35] Ba is less hydrated than Ca, so the hydration factor does not limit Ba incorporation; however, Ca and Ba also have large differences in ionic size (Ca, 1.00 Å; Ba, 1.35 Å).[32] Nevertheless, we suggest that the calcite lattice, by losing its rotational order, can 'accommodate' itself to the lattice distortions caused by Ba, and therefore does not limit Ba incorporation.

### 2.3. Thermal Behavior

In-situ annealing was performed here on fully disordered Ba-substituted calcite (balcite) and on partially ordered Ba-substituted calcite (Ba-calcite) powders. These two phases behaved differently. After being heated at 200 °C, Ba-calcite almost completely decomposed into two phases (**Figure 4**): an $R\bar{3}m$ Ba-enriched balcite phase (with peaks shifted to lower angles and the absence of (113) peak) and a more ordered Ba-depleted calcite. phase (with peaks shifted to higher angles and a higher intensity (113) peak). Therefore, Ba-substituted calcite with partial carbonate group orientation order is a metastable phase and transforms into the fully ordered and fully disordered states at elevated temperatures.

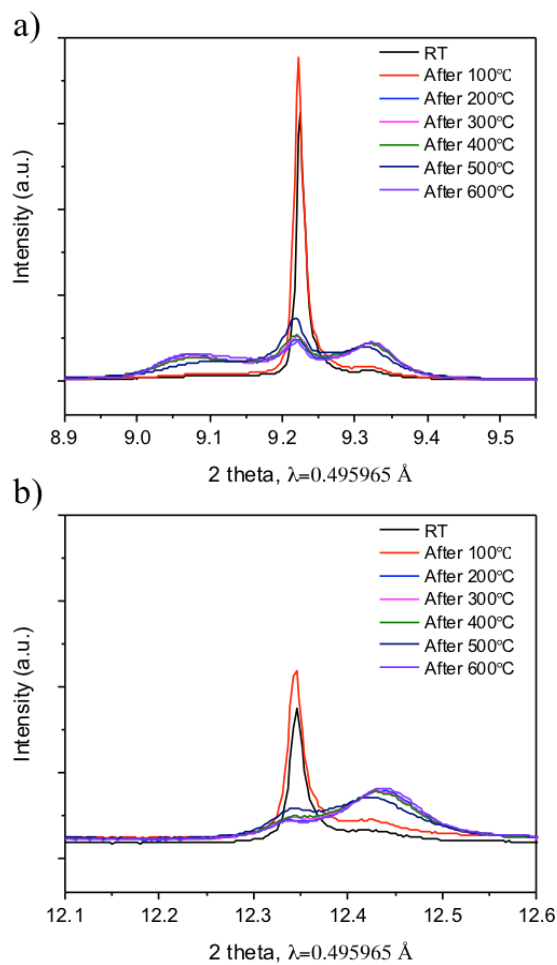

**Figure 4**. (a) (104) and (b) (113) calcite reflections obtained from partially ordered Ba-calcite powders prepared via the hydrothermal synthesis route from solutions containing 20% Ba and in-situ annealed at 100, 200, 300, 400, 500, 600, and 700 °C (Data acquired from a synchrotron source, λ=0.49597 Å.)

Balcite, however, was stable up to 700°C (**Figure 5**) and underwent only slight changes in lattice parameters during annealing: the a-parameter increased and the c-parameter decreased before increasing again after heating at ≥400°C (Figure 5b). This finding suggested that balcite lattice rearranges itself; an explanation, based on our model, is given later in the article. Results are shown for powders synthesized using solutions containing 20% and 30% Ba, but Ba-calcite (partially ordered) decomposition and balcite (disordered) stability were observed also for powders with other compositions (data not shown). These results strongly indicated, that the disordered balcite is a low-temperature stable phase whereas the partially disordered Ba-calcite phase is metastable and can decompose to disordered balcite and ordered calcite.

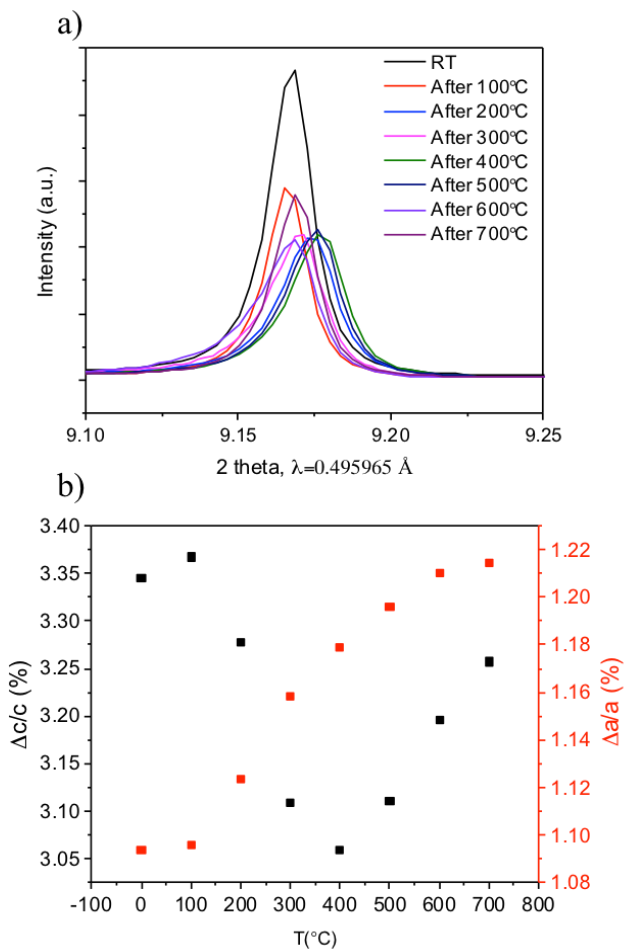

**Figure 5**. (a) (104) calcite reflections obtained from disordered balcite powders prepared via the hydrothermal synthe-sis route from solutions containing 30% Ba and in-situ an-nealed at 100, 200, 300, 400, 500, 600, and 700 °C. (Data acquired from a synchrotron source, λ=0.39999 Å, and rescaled at λ=0.49597 Å.) (b) Lattice distortions relative to pure calcite (Δc/c in black, Δa/a in red) as a function of annealing temperature.

### 2.4. $R\bar{3}c$ to $R\bar{3}m$ Phase Transition – model

We developed a quasichemical model describing the $R\bar{3}c$ to $R\bar{3}m$ phase transition (see Supplementary Information) which considers the mixture of different anionic group orientations as a binary mixture of different "atoms", A and B, with different pair bonds. According to this model, the temperature of transition from the ordered to disordered state is proportional to the free energy change when a single anionic group is inverted (rotated through 60º) in an otherwise perfectly ordered lattice. This free energy change depends on the oxygen-oxygen interaction energy, and thus, on the interatomic distances in calcite or in Ba-calcite. Using the known oxygen-oxygen interaction potential in calcite[36] and having in mind the determined lattice distortions caused by the substitution of Ca by Ba, we were able to compare the calculated order-disorder transition temperatures in pure calcite and Ba-calcites with different Ba concentrations.

According to our refinements, Ba incorporation into calcite causes an increase in the oxygen-oxygen and a decrease in the oxygen-carbon distances. In this model, we consider the interactions between every 3 oxygen atoms of one anionic group (the central group in the **Figure 6**) and the oxygen atoms of all the first-neighbour anionic groups, both in the plane of the central group and out of the plane (located in the adjacent planes). It should be noted that interatomic Ca (or Ba) – oxygen interactions are unchanged by the rotation of carbonate anionic groups for 60°, while oxygen-oxygen interactions are changed and therefore only the latter are considered in the model. The energy differences between the ordered and disordered states are calculated considering the 2 different orientations presented in Figure 6: the ordered state (Figure 6a) and the disordered state with the central anionic group rotated for 60° (Figure 6b).

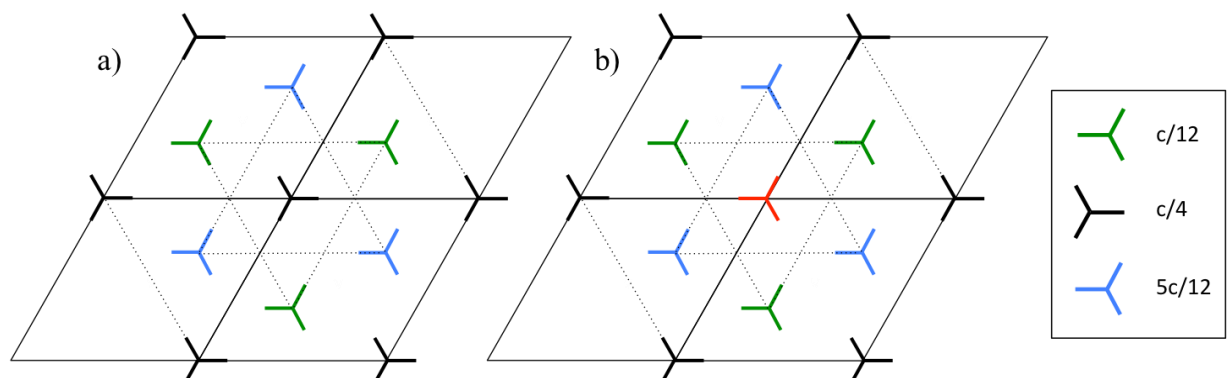

**Figure 6**. Carbonate groups' positions from the planes $c/12$, $c/4$ and $5c/12$ projected onto the plane perpendicular to the c-axis. a) All the carbonate groups are ordered. b) All the carbonate groups are ordered except for the middle one (in red) that is in opposite orientation (rotated by 60°).

For each position, 108 interactions are considered: between the 3 oxygen atoms of the anionic group in the centre with the oxygen atoms from the 6 closest groups in the plane and 6 closest out-of-plane anionic groups. The oxygen-oxygen distances are calculated using reported values of lattice parameters, a, c and x for pure calcite at the order-disorder transition temperature,[30] and the lattice parameters of the Ba-calcite obtained from the Rietveld refinements. The oxygen-oxygen interaction energies in calcite are calculated using the Buckingham potential:[36-37]

$$\Phi(r_{ij}) = A_0 \exp\left(-\frac{r_{ij}}{\rho_0}\right) - \frac{c_0}{r_{ij}^6} \quad (1)$$

where $r_{ij}$ are the distance between oxygen atoms, and the constants $A_0$, $\rho_0$ and $c_0$ for oxygen-oxygen interaction in calcite are from Pavers et al.[36] $A_0$=16732.0eV, $\rho_0$=0.213 Å, $c_0$=3.47eV·Å$^6$. The energy differences between the disordered and ordered states were calculated as following:

$$\omega = z\Delta\varepsilon_0 = \sum_{i=1}^{3}\sum_{j=1}^{36}[\Phi(r_{ij}^{disordered}) - \Phi(r_{ij}^{ordered})] \quad (2)$$

where z is the coordination number (z=6 in the plane of the central anionic group, where disordering is considered), i=1,2,3 correspond to oxygen atoms of the central carbonate group, while j=1…36 to oxygen atoms in all nearest groups. The calculated energies and corresponding transition temperatures are presented in **Figure 7**a and Tables S2,S3 in Supplementary Materials. The transition temperatures for the Ba containing calcites were calculated using the proportion:

$$\frac{T_c^{Bal}}{T_c^{Cal}} = \frac{\omega^{Bal}}{\omega^{Cal}} \quad (3)$$

where $\omega^{cal}$ and $\omega^{Bal}$ are the differences of oxygen-oxygen interaction energies, eq.(2), for calcite and balcite, respectively, $T_c^{Cal}$=1240K is the order-disorder transition temperature in pure calcite. As can be seen, the calculated energy difference $z\Delta\varepsilon_o$ for pure calcite corresponds very well to the known transition temperature: for z=6 the calculated transition temperature T'$_c$=$\Delta\varepsilon_0$/k$_B$=1172K (the discrepancy with $T_c^{Cal}$=1240K can be mainly related to neglected vibrational entropy terms). With increase of Ba concentration the transition temperature is substantially reduced, from 1240 K for pure calcite, to 362 K for the balcite containing 62% Ba (Figure 7a, Table S2). Room temperature (as it was expected from our experimental results) is not reached in these model calculations; however, the substantial decrease of the transition temperature (more than 2-3 times as compared to pure calcite) indicates that the increase of interatomic distances upon Ba incorporation is indeed a crucial factor for calcite disordering. In addition, one can assume that parameters of the oxygen-oxygen potential[36-37] in the presence of substantial amount of Ba could be slightly changed. If, for example, the radius of repulsive interaction, $\rho_0$, in the presence of large amount of substituted Ba would decrease for only 5%, the calculated transition temperature of the calcite with 38%Ba decreases to 310 K (Figure 7a, Table S4). Therefore, increase of oxygen-oxygen distances induced by incorporation of Ba in the calcite lattice results in a smaller energy gap between the ordered and disordered states that may lead to stable disordering of balcite at much lower temperatures as compared to calcite.

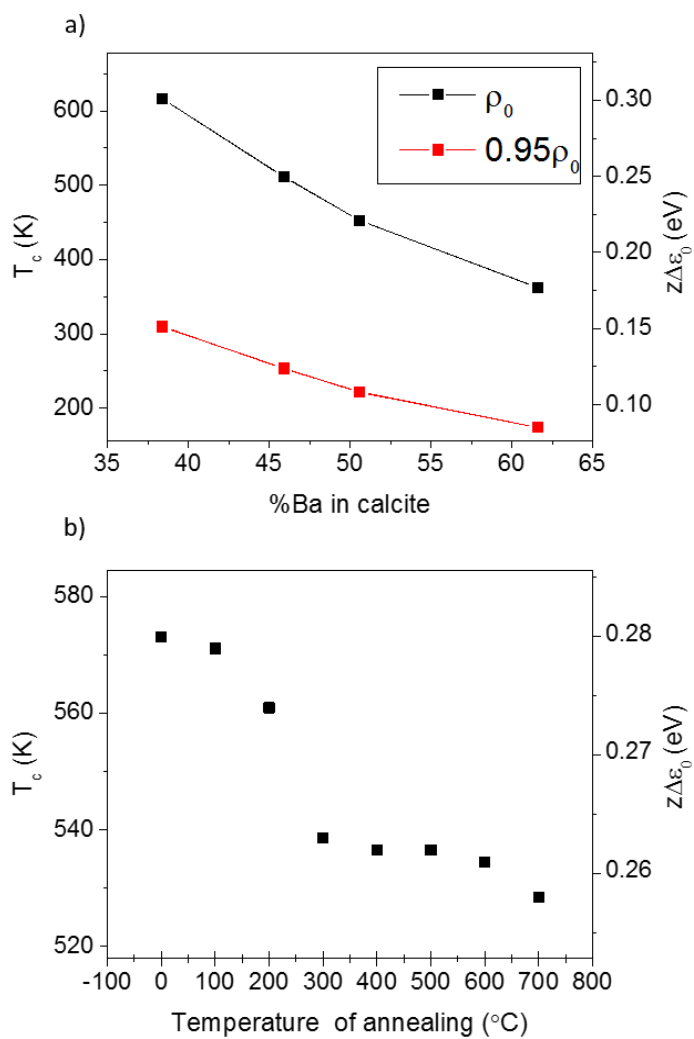

**Figure 7**. Calculated transition temperatures (left axis) and energy differences between the disordered and ordered states (right axis) for: a) balcite powders, synthesized via hydrothermal synthesis with different Ba concentrations; parameters of the oxygen-oxygen Buckingham potential for calcite37, 38 were used (black), eq.(1), and the same parameters with a reduced

repulsive interaction radius (red); b) balcite powder, synthesized via hydrothermal synthesis, using a 30% Ba-solution, after annealing at 100, 200, 300, 400, 500, 600, and 700°C.

These calculations also help us to understand the lattice parameter minor changes observed upon the in-situ annealing of balcite (Figure 5b). Indeed, calculations of $\Delta\varepsilon_0$ using the post-annealing lattice parameters of balcite showed that it slightly decreased after annealing (Figure 7b, Table S3). Correspondingly, the temperature of transition decreased as well, from 573 K (before annealing) to 528 K after annealing at 700°C. Therefore, the annealing actually allowed the relaxation of the balcite lattice towards a more stable disordered state.

## 3. Conclusion

Substitution of Ca by large amounts of Ba in calcite results in disordered balcite phases, having $R\bar{3}m$ symmetry. We studied in detail the transition caused by Ca substitution from the ordered to the disordered phase, using powders prepared via hydrothermal synthesis from a Ba-ACC precursor and over a wide range of Ba content. The incorporation of Ba into the calcite lattice caused an expansion of the lattice cell and progressive disorder in carbonate orientations until their full randomness. We showed that in all the synthesis routes tested, this disordered calcite phase forms if the solutions contain more than 30% Ba. With 20% Ba solutions, only the partially ordered Ba-substituted calcite was formed by fast-rate syntheses while the fully disordered balcite could be formed by slow-rate synthesis. Moreover, disordered balcite was stable at high temperatures whereas the partially ordered Ba-calcite decomposed to $R\bar{3}m$ balcite and almost pure calcite after being heated at only 200 °C. Taken together, our results indicated that the $R\bar{3}c$ Ba-calcite is metastable while the $R\bar{3}m$ balcite is the stable phase which forms in the presence of sufficient amount of Ba.

## 4. Experimental Section

*Synthesis Routes.* We studied the incorporation of Ba into calcite by using different synthesis routes: hydrothermal synthesis, vapor diffusion synthesis, and a drop-by-drop addition technique.

*Hydrothermal synthesis.* For this synthesis we used Ba-ACC as the amorphous precursor, which we synthesized by mixing a solution consisting of 200 mM [$BaCl_2$]+[$CaCl_2$], with a 200-mM solution of $Na_2CO_3$. The mixed solution was placed in an autoclave under conditions of high temperature and pressure. After 2 h at 135 °C in the autoclave the powder was washed, filtered and dried.

*Vapor diffusion synthesis.* Vials containing solutions consisting of 10 mM [$BaCl_2$]+[$CaCl_2$] were placed in a sealed desiccator in the presence of $CO_2$ gas from ammonium bicarbonate powder. The vials were covered with parafilm punctured by three holes in order to minimize the contact area and slow down the reaction. After 4 days the powder was washed, filtered and dried.

*Drop-by-drop addition synthesis.* Solutions consisting of 10 mM [$BaCl_2$]+[$CaCl_2$] were added to a solution of $Na_2CO_3$ at a controlled rate of 0.05 mL/min using a peristaltic pump.

*X-Ray Diffraction (XRD).* Some of the diffraction patterns were acquired at the ID22 powder diffraction beamline at the European Synchrotron Research Facility (ESRF), Grenoble, France. For these experiments we utilized synchrotron light at wavelengths of 0.39999 Å and 0.4959652 Å.

*Data Analysis.* Diffraction data were analysed using Rietveld refinements with GSAS-II software.[38] The amounts of Ba that substituted for Ca in calcite were quantified by refining the occupancy sites. Such refinement was reliable in this system owing to the marked difference in the structure factors of Ca and Ba. All refinements had a goodness of fit under 12 and $\chi^2$ under 14.

*Inductively Coupled Plasma (ICP).* ICP was performed on solutions containing 3 vol% of HCl and a few ppm of dissolved Ba-calcite.

*High-Resolution Scanning Electron Microscope (HRSEM).* Images of carbon-coated samples were taken by using the Zeiss-Ultra Plus FEG-SEM.

*Scanning Transmission Electron Microscope – Energy Dispersive Spectroscopy (STEM-EDS).* The Titan Themis G$^2$ 300 was used to investigate a thin cross-section, which was prepared using Focused Ion Beam (FIB) technology.

**Supporting Information**
Supporting Information is available from the Wiley Online Library or from the author.

**Acknowledgements**
This project has partially received funding from the European Union Horizon 2020 research and innovation program under the Marie Skłodowska-Curie grant agreement no. 642976-NanoHeal Project. We are grateful to Shir Alpert, Achiya Livne, Sylwia Mijowska and Arad Lang for their help collecting data at the ESRF.

Received: ((will be filled in by the editorial staff))
Revised: ((will be filled in by the editorial staff))
Published online: ((will be filled in by the editorial staff))

Supporting Information

**Formation and Stability of Ordered and Disordered Ba-substituted Calcite Phases**

*Eva Seknazi, Davide Levy†, Iryna Polishchuk, Alex Katsman, Boaz Pokroy\**

**Table S1**. Goodness of fit and lattice parameters of the main phase obtained from hydrothermal syntheses, extracted from our refinements.

| %Ba in solution | GOF | wR % | Phase | Ba fraction | sig | a (Å) | esds | C (Å) | esds | x | sig |
|---|---|---|---|---|---|---|---|---|---|---|---|
| 10 | 7.3 | 12.77 | R-3c | 0.225 | 0.01 | 5.00533 | 0.00009 | 17.31813 | 0.0003 | 0.2515 | 0.001 |
| 15 | 11.54 | 14.18 | R-3c | 0.30 | 0.01 | 5.01542 | 0.00008 | 17.42898 | 0.0003 | 0.2473 | 0.001 |
| 20 | 9.89 | 12.08 | R-3c | 0.359 | 0.009 | 5.02449 | 0.00006 | 17.51584 | 0.0002 | 0.2412 | 0.0009 |
| 25 | 7.36 | 9.96 | R-3c | 0.40 | 0.02 | 5.03153 | 0.00016 | 17.55559 | 0.0006 | 0.2434 | 0.001 |
|  |  |  | R-3m | 0.33 | 0.01 | 5.03513 | 0.00006 | 8.79199 | 0.00002 | 0.2485 | 0.001 |
| 30 | 5.61 | 8.93 | R-3m | 0.384 | 0.008 | 5.04622 | 0.00005 | 8.82785 | 0.00007 | 0.2473 | 0.0008 |
| 35 | 7.66 | 9.59 | R-3m | 0.459 | 0.009 | 5.063 | 0.00006 | 8.88 | 0.0009 | 0.2441 | 0.008 |
| 40 | 7.99 | 9.26 | R-3m | 0.506 | 0.009 | 5.07986 | 0.00007 | 8.92769 | 0.0001 | 0.2426 | 0.0008 |
| 50 | 6.46 | 9.53 | R-3m | 0.616 | 0.01 | 5.10999 | 0.0001 | 9.02730 | 0.0002 | 0.2398 | 0.0008 |

**1. Short-Range Ordering of CO$_3$ groups in Calcite**

The alternate orientation of CO$_3$ groups in alternate layers of the calcite crystalline structure can be considered as a short range ordering. Following to the quasichemical model for short ordering developed by Guggenheim and Fowler[1,2] one can describe the mixture of different orientations as a binary mixture of different "atoms", A and B, with different pair bonds, $\varepsilon_{AA}, \varepsilon_{BB}$ and $\varepsilon_{AB}$, A and B correspond to different orientations of an anionic group (rotated for 60° to each other).

Let us consider three parallel atomic planes of anionic CO$_3$ groups, and consider different possible orientation of these groups in the central plane assuming that orientations in two adjacent planes are ordered. Let us now assume that the number of neighbouring A- groups with the same orientation in x-direction equals to N$_{AA}$, with the same orientation in the (-x)-direction is $N_{BB}$, and the number of pairs with opposite orientation equals $N_{AB}$. The total number of $CO_3$ groups with orientation in x-direction is $N_A$, and in the (–x)-direction is $N_B$. The mass balance equations are:

$$N_A = N_{AA} + \frac{N_{AB}}{2} \text{ and } N_B = N_{BB} + \frac{N_{AB}}{2} \qquad (1S)$$

The mixing enthalpy of this binary structure:

$$\Delta h = \frac{1}{2}N_0 z(X_{AA}\varepsilon_{AA} + X_{BB}\varepsilon_{BB} + X_{AB}\varepsilon_{AB} - X_A\varepsilon_{AA} - X_B\varepsilon_{BB}) = \frac{1}{2}N_0 z\Delta\varepsilon_0 X_{AB} = \omega X_{AB}/2 \quad (2S)$$

where $X_{ij} = 2N_{ij}/N_0 z$, $X_i = 2N_i/N_0 z$, $i,j = A,B$, $\Delta\varepsilon_0 = \varepsilon_{AB} - (\varepsilon_{AA} + \varepsilon_{BB})/2$. It should be noted that, in general case, $\varepsilon_{AA} \neq \varepsilon_{BB}$, since these effective pair energies should include interactions with anionic groups from parallel planes, and they are different for different orientations of "atoms"; $N_0$ is the Avogadro's number, z is the coordination number. The mixing entropy[1]:

$$\Delta s_m = -R(X_A \ln X_A + X_B \ln X_B) - \frac{Rz}{2}\left(X_{AA}\ln\frac{X_{AA}}{X_A^2} + X_{BB}\ln\frac{X_{BB}}{X_B^2} + X_{AB}\ln\frac{X_{AB}}{2X_A X_B}\right) + \frac{1}{2}\eta X_{AB} \quad (3S)$$

where the first term corresponds to configurational mixing entropy of random distribution in $CO_3$ group orientations, the second term takes into account possible deviation from the random distribution and the last term is the non-configurational mixing entropy.

The value of $X_{AB}$ at any composition is that which minimizes the Gibbs mixing energy at constant numbers $X_A$ and $X_B$:

$$\left.\frac{\partial(\Delta h_m - T\Delta s_m)}{\partial X_{AB}}\right|_{X_A, X_B} = 0 \qquad (4S)$$

Since orientations A and B can transform to each other, the Gibbs mixing energy should be also minimized with respect to this transformation:

$$\left.\frac{d(\Delta h_m - T\Delta s_m)}{dX_A}\right|_{X_A + X_B = 1} = 0 \qquad (5S)$$

Using eqs. (2S)-(4S) one can obtain the well-known expression:

$$\frac{X_{AB}^2}{X_{AA}X_{BB}} = 4A^2 \qquad (6S)$$

where $A \equiv \exp[-\frac{\omega-\eta T}{zRT}]$. The interaction energy between equally oriented neighbouring $CO_3$ groups is stronger than between oppositely oriented ones, that means $\Delta\varepsilon_0 > 0$ and $\omega > 0$. At low temperatures, when $A \ll 1$, the concentration of pairs $X_{AB} \ll 1$, and almost all neighbouring $CO_3$ groups are equally oriented in a given plane. From eqs.(1S),(6S):

$$X_{AB} = \frac{A^2}{1-A^2}[\sqrt{1+\frac{4X_A X_B(1-A^2)}{A^2}} - 1 \tag{7S}$$

and for low temperatures ($A \ll 1$), $X_{AB} \approx 2A\sqrt{X_A X_B}$.

Order-disorder transition temperature is usually defined from (6S) as

$$T_c = \frac{w}{zR+\eta} \approx \frac{\Delta\varepsilon_0}{k_B} \tag{8S}$$

where $k_B$ is the Boltzmann constant. One can note that the proportionality between critical temperature of order-disorder transition in calcite and the energy $w$ has been obtained by Antao et al.[3] in their mean-field Bragg-Williams approximation model, where they defined this temperature as $T_c = w/2k_B$, which, however, did not take into account the real crystal structure of the calcite.

Below the critical temperature, a certain amount of groups with "wrong" orientation may exist. In order to find their amount, we should use the second condition of free energy minimum, (5S), from which the following approximated equation can be obtain:

$$\sqrt{y}\ln y; A\ln A \tag{9S}$$

where $y = X_B/X_A$. Numerical solution of eq.(9S) is presented in Fig.1S. As can be see, the number of "wrong" orientations is small below critical temperature, but still exists and almost constant in the range $(0.8 \div 1.0)T_c$. This may explain an existence of a partially ordered phase IV in calcite and in Ba-calcites.

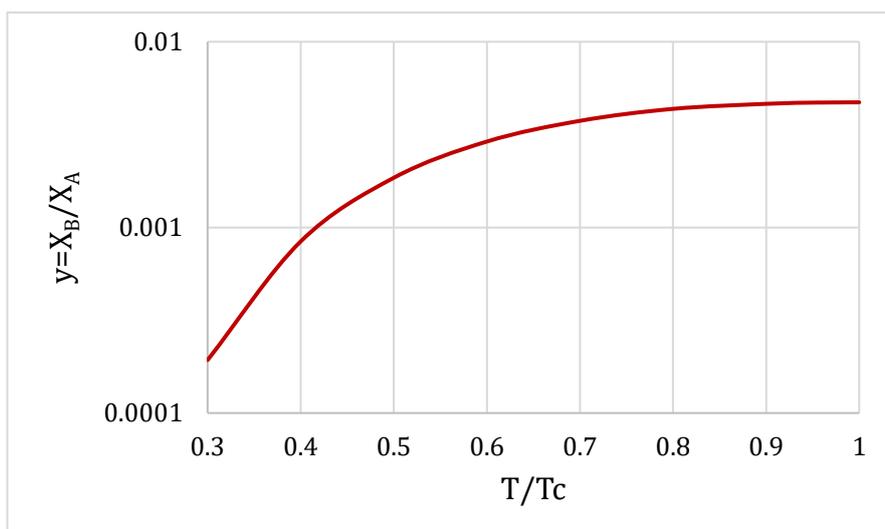

**Fig. S1**. The ratio of oppositely oriented anionic groups as a function of temperature, calculated from eq.(9S).

## 2. Calculated values of oxygen-oxygen interaction energies and order-disorder transition temperatures in the Ba-calcite phases

The oxygen-oxygen interaction energies for ordered state (Figure 6a) and disordered state (Figure 6b) were calculated according to eq. (2) by using oxygen-oxygen interatomic potential, eq. (1). Order-disorder transition temperatures were estimated using the proportion, eq. (3), presented in the main text of the paper.

**Table S2.** Calculated oxygen-oxygen interaction energies in the ordered and disordered state, their difference and the order-disorder transition temperature.

| %Ba in solution | %Ba in calcite | $z\varepsilon_{ordered}$ (eV) | $z\varepsilon_{disorder}$ (eV) | $z\Delta\varepsilon$ (eV) | $T_c$ (K) |
|---|---|---|---|---|---|
| 0 | 0 | -0.008 | 0.598 | 0.606 | 1240 |
| 30 | 38.4 | -0.021 | 0.279 | 0.301 | 616 |
| 35 | 45.9 | -0.025 | 0.225 | 0.25 | 511 |
| 40 | 50.6 | -0.027 | 0.194 | 0.221 | 453 |
| 50 | 61.6 | -0.031 | 0.146 | 0.177 | 362 |

**Table S3**. The calculated order-disorder transition temperatures with oxygen-oxygen interatomic Buckingham potential[36,37], eq. (1), but $\rho_0^{'} = 0.95\rho_0$

| Ba% | $\omega^{Bal}/\omega^{cal}$ | $z\Delta\varepsilon$, eV | $z\varepsilon_{disorder}$, eV | $z\varepsilon_{ordered}$, eV | $T_c$, K |
|---|---|---|---|---|---|
| 30 | 0.25 | 0.151 | 0.094 | -0.057 | 310 |
| 35 | 0.204 | 0.124 | 0.067 | -0.057 | 253 |
| 40 | 0.179 | 0.108 | 0.052 | -0.056 | 222 |
| 50 | 0.141 | 0.085 | 0.029 | -0.056 | 174 |

**Table S4**. The calculated oxygen-oxygen interaction energy differences between the disordered and ordered states and transition temperatures in 30%Ba-calcite after heating

| $T_{heat}$, °C | $\omega^{Bal}/\omega^{cal}$ | $z\Delta\varepsilon_{disorder}$, eV | $T_c$, K |
|---|---|---|---|
| 0 | 0.462 | 0.280 | 573 |
| 100 | 0.460 | 0.279 | 570 |
| 200 | 0.453 | 0.274 | 561 |
| 300 | 0.434 | 0.263 | 538 |
| 400 | 0.433 | 0.262 | 537 |
| 500 | 0.432 | 0.262 | 536 |
| 600 | 0.431 | 0.261 | 534 |
| 700 | 0.426 | 0.258 | 529 |

The values $z\varepsilon_{\text{ordered}}$ and $z\varepsilon_{\text{disordered}}$ are the oxygen-oxygen interaction energies defined as follows:

$$z\varepsilon_{\text{ordered}} = \sum_{i=1}^{3}\sum_{j=1}^{36}[\Phi\left(r_{ij}^{ordered}\right)], \; z\varepsilon_{\text{disordered}} = \sum_{i=1}^{3}\sum_{j=1}^{36}[\Phi\left(r_{ij}^{disordered}\right)] \qquad (10S)$$

where parameters of the interaction potential $\Phi(r_{ij})$ are given in the main text of the paper.